\documentclass[a4paper,fleqn,usenatbib]{mnras}

\usepackage{newtxtext,newtxmath}
\usepackage[T1]{fontenc}
\usepackage{ae,aecompl}
\usepackage{graphicx}
\usepackage{amsmath}
\usepackage{amstext}
\usepackage{multirow}

\title[Synchrotron emitting Komissarov torus with magnetic polarization around Kerr black holes: ]{Synchrotron emitting Komissarov torus with magnetic polarization around Kerr black holes}

\author[J. M. Velásquez-Cadavid et al.]{J. M. Velásquez-Cadavid$^{1}$\thanks{juan2208056@correo.uis.edu.co}, Fabio D. Lora-Clavijo$^{1}$\thanks{fadulora@uis.edu.co}, Oscar M. Pimentel$^{1}$\thanks{oscar.pimentel@correo.uis.edu.co}, 
\newauthor
and J. A. Arrieta-Villamizar$^{1}$\thanks{jesus2208058@correo.uis.edu.co}
\\
$^{1}$Grupo de Investigaci\'on en Relatividad y Gravitaci\'on, Escuela de F\'isica, Universidad Industrial de Santander, A. A. 678, Bucaramanga 680002, Colombia
}

\date{Accepted XXX. Received YYY; in original form ZZZ}

\pubyear{2021}

\begin{document}
\label{firstpage}
\pagerange{\pageref{firstpage}--\pageref{lastpage}}
\maketitle

\begin{abstract}

Magnetic fields in black hole accretion disks are associated with processes of mass accretion and energy amplification. The contribution of the magnetic field due to the magnetic polarization of the material induces effects on the physical properties of the medium that have repercussions on the radiation coming from the accretion disks. Hence, from observations, it could be possible to infer the "fingerprint" left by the magnetic polarization of the material and establish the properties of the spacetime itself. As the first step in this purpose, we use numerical simulations to systematically analyze the possible observable effects produced by the magnetic properties of an accretion disk around a Kerr black hole. We found that under the synchrotron radiation power-law model the effects of the magnetic polarization are negligible when the plasma is gas pressure-dominated. Nevertheless, as beta-plasma decreases, the emission becomes more intense for magnetic pressure-dominated disks. In particular, we found that paramagnetic disks emit the highest intensity value independent of the beta-plasma parameter in this regime. By contrast, the emitted flux decreases with the increase of beta-plasma due to the dependence of the magnetic field on the emission and absorption coefficients. Moreover, the disk morphology changes with the magnetic susceptibility: paramagnetic disks are more compact than diamagnetic ones. This fact leads to diamagnetic disks emitting a greater flux because each photon has a more optical path to travel inside the disk.

\end{abstract}


\section{Introduction}
\label{sec:intro}

General relativity is nowadays the standard physical theory to describe gravitational phenomena. From a theoretical and astrophysical point of view, black holes are one of the most successful predictions of general relativity, testing this theory under regimes of extreme gravity. This is the case of the gravitational waves produced by the coalescence of two black holes \citep{gw_merger}, the EHT observations inside the galaxy M87 \citep{eht}, and the very recent observation of Sagittarius A* \citep{SgrA1, SgrA2, SgrA3, SgrA4, SgrA5, SgrA6}. Those observations were possible by detecting radiation from the accretion disk around the black hole. Photons that pass marginally near the event horizon and escape from the gravitational pull make up the shadow of the black hole \citep{Falcke_2000}. That radiation is generated either because the cloud of gas and dust in the accretion disk heats up \citep{thermal} or due to the acceleration of charges inside the disk \citep{non-thermal}.

Accretion disks are intimately linked to black holes since it is possible to extract information both from space-time itself and from the matter that makes up the disk, allowing to describe of astrophysical scenarios such as active nuclei of galaxies or microquasars \citep{living-review}. Thus, the study of accretion disks has been a focus of interest for several decades, so it is possible to find in the literature some analytical solutions to model equilibrium accretion disks around black holes \citep{Kozlowski, Fragile_polish, Pugliese}. In particular, when we talk about accretion disks, the study of the magnetic field stands out since that could be responsible for the generation of instabilities in the disk that would give rise to the accretion process \citep{magneto-rotational}. Besides, it could accelerate and collimate the gas in relativistic jets that emanate towards the interstellar medium from the accretion disk \citep{blandford}. Based on this, some models describe magnetized tori around black holes \citep{okada1989model,komissarov, font_tori, Fragile_magnetized, Soler},  being Komissarov's work particularly noteworthy, where he shows for the first time an analytical solution to describe tori in magnetohydrodynamic equilibrium around Kerr black holes. From that result, it is possible to show that magnetized disks are unstable to non-axial disturbances, in addition to the fact that energy dissipation and angular momentum transport are a consequence of magnetorotational instability \citep{fragile2}. 

An interesting aspect to take into account is the degree of magnetic polarization of the material and its effects on the dynamics of the disk. \cite{Oscar} (Pimentel, from now on) established, based on Komissarov's work, a model to describe magnetized tori with arbitrary magnetic susceptibility around Kerr black holes. In this paper, they found that magnetic susceptibility affects the compactness of the disk, becoming more compact for paramagnetic disks and less compact for diamagnetic disks. Additionally, the impact of the magnetic susceptibility, on the development of the magneto-rotational instability in weakly magnetized accretion disks, was studied in \citep{Pimentel_2021}. This work shows that paramagnetic disks have larger turbulent structures than diamagnetic ones. Moreover, the beta-plasma parameter shows that the paramagnetic disk becomes more strongly magnetized than the no polarized case, which itself is more strongly magnetized than the diamagnetic disk. As it can be seen, the inclusion of the magnetic properties of the material in the dynamics of the relativistic accretion disks is relevant to understanding the different physical processes happening in this system. It is worth mentioning that magnetic polarization has been considered in other astrophysical scenarios such as neutron stars \citep{blandford1982magnetic,suh2010magnetic,chatterjee2015consistent,wang2016diamagnetic}, relativistic waves \citep{Pimentel_2018}, the Kelvin-Helmholtz instability in relativistic plasmas \citep{Pimentel_2019}, among others.  

Inspired by the above, an interest arises in determining the potential impact that magnetic polarization may have on future observations. For this reason, we study the geodesics of the photons emanating from the torus and that manage to escape the gravitational attraction of the black hole. That is where numerical simulations play a relevant role. Nowadays, ray tracing codes represent an essential tool for testing some models of accretion disks around black holes when comparing observations. In the literature, it is possible to find codes such as \texttt{BHAC} \citep{bhac}, \texttt{RAPTOR} \citep{raptor}, \texttt{HARMRAD} \citep{harm}, among others characterized by their computing power and the ability to simulate more realistic astrophysical scenarios by numerically solving the equations of relativistic magnetohydrodynamics with radiative terms. To determine the impact of the magnetic polarization on the disk radiation intensity map, we use \texttt{OSIRIS} \citep{OSIRIS}, a code of our authorship that is based on the Hamiltonian formalism and solves null geodesics through the tracing inverse of rays. This code has been validated by successfully reproducing thin accretion disks around Kerr black holes, simulating the shadow of compact objects with arbitrary quadrupole and naked singularities, as well as reproducing time-like orbits around these exotic bodies \citep{Arrieta_Villamizar_2020}. 

Thus, in this paper, we carry out for the first time simulations of the intensity map and the emission-line profiles for a magnetized torus considering the magnetic polarization of the fluid. We found that magnetic susceptibility modifies the intensity and the emission-line profiles depending on how magnetized it is the disk, by varying the magnetic pressure and showing in the map of intensities the changes in the compactness of the torus. The organization of this article is as follows: Section 2 We make a brief description of the magnetically polarized disk model developed by Pimentel, where a toroidal magnetic field is assumed and the particles that make up the torus move in circular orbits; Section 3 describes the radiative transport theory and the radiation mechanism employed; Section 4 briefly describes the code used for the simulations and the corresponding numerical configurations; finally, in Section 5 we show the results of our simulations of magnetically polarized tori around a Kerr black hole. Throughout this document, we will work with a signature $(-,+,+,+)$ and geometrized units, where $G=c=1$, being $G$ the gravitational constant, and $c$ the speed of light. Paramagnetic materials present a magnetic susceptibility $\chi > 0$, and diamagnetic materials, $\chi < 0$.

\section{KOMISSAROV TORUS WITH MAGNETIC POLARIZATION IN KERR SPACE-TIME}
\subsection{Magnetic susceptibility}

For an electron gas, the magnetic polarization is composed of the paramagnetic and diamagnetic contribution \citep{magnetism_landau}. Paramagnetism is associated with the intrinsic magnetic moment of electrons and the alignment of the spins due to magnetic torques, while diamagnetism is associated with the electron orbital motion around the magnetic field lines. In particular, the quantization of the orbital angular momentum of free electrons in a gas around magnetic field lines. For free atoms, the magnetic susceptibility is written as follows \citep{weber}
\begin{equation}
    \chi_m = -\frac{e^2\mu_0}{6m_e}\sum_Z\langle R^2 \rangle, \label{eq:chi_diam}
\end{equation}
where $\mu_0$ is the magnetic permeability in the vacuum, $e$ and $m_e$ are the charge and mass of the electron, respectively, $Z$ is the number of atoms, and $\langle R \rangle$ is the average radii of electrons orbiting around the nucleus assuming spherical symmetry. On the other hand, for paramagnetic materials, the magnetization is calculated as 
\begin{equation}
    M = ng\mu_{\text{B}}JB_{j}(T),
\end{equation}
with $n$ the density of particles, $\mu_{\text{B}}$ the Bohr magneton and  $g$ the Landé factor,
\begin{equation}
g = 1+\frac{J(J+1)+S(S+1)-L(L+1)}{2 J(J+1)},
\end{equation}
being $J$, $S$ and $L$ the total, spin, and orbital magnetic quantum number, respectively. $B_j(T)$ corresponds to the Brillouin function, 
\begin{equation}
B_J(\alpha)=\frac{2 J+1}{2 J} \operatorname{coth}\left(\frac{(2 J+1) \alpha}{2 J}\right)-\frac{1}{2 J} \operatorname{coth}\left(\frac{\alpha}{2 J}\right), 
\end{equation}
where $T$ is the temperature, and
\begin{equation}
\alpha=\frac{g \mu_0 \mu_{\mathrm{B}} J H}{k_{\mathrm{B}} T}, \label{eq:alpha}
\end{equation}
with $H$ the magnetic field intensity. Thus, the magnetic susceptibility is calculated as $\chi_m = \partial M/\partial H$.

\begin{figure}
\centering
   \includegraphics[width=8.5cm]{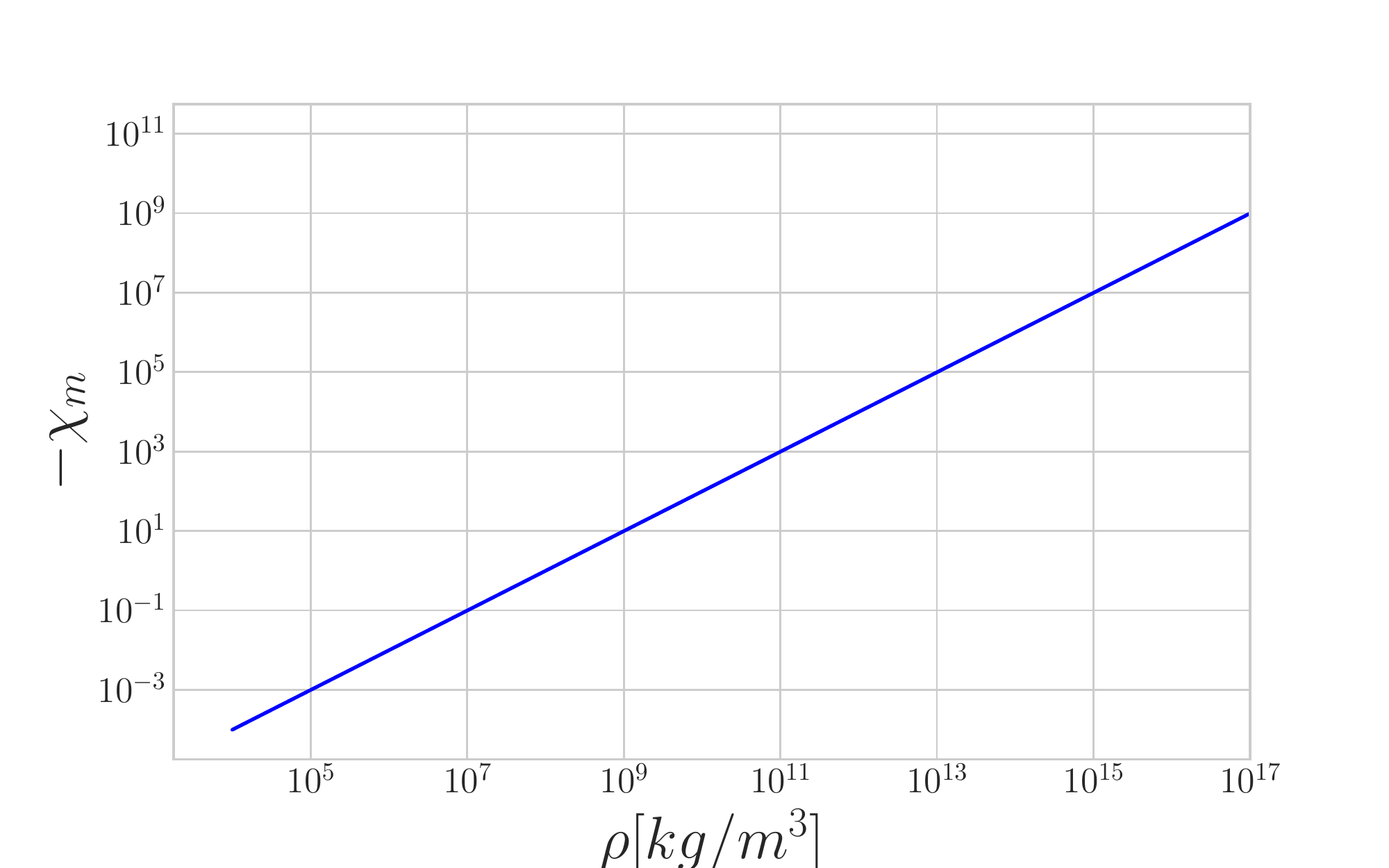}
    \caption{Magnetic susceptibility for diamagnetism as a function of density, in logarithmic scale, for a hydrogen plasma. Magnetic susceptibility increases as density increase too, showing a linear dependence. Values of density were taken considering from an accretion disk to a neutron star.}
    \label{fig:chi_rho}
\end{figure}

For diamagnetic materials, magnetic susceptibility only depends on the number of particles. Considering a hydrogen gas, we calculate the susceptibility in a range from densities in accretion disks ($10^{5} [kg/m^3] - 10^{7} [kg/m^3]$, \cite{cold_plasma})  to densities in neutron star densities ($10^{19} [kg/m^3]$, \cite{neutron-star}). In the model we are using, the magnetic susceptibility looks proportional to density, as we are showing in figure \eqref{fig:chi_rho}. 
For accretion disk, the scales for magnetic susceptibility are of the order of $10^{-3} - 10^{-1}$, while for more compact objects like neutron stars, these scales are of the order of $10^{9} - 10^{11}$. However, it is unknown if this model applies to such high densities. Nevertheless, it is the first approach to determine the magnetic susceptibility of different diamagnetic astrophysical objects.

\begin{figure*}
\centering
   \includegraphics[width=17cm]{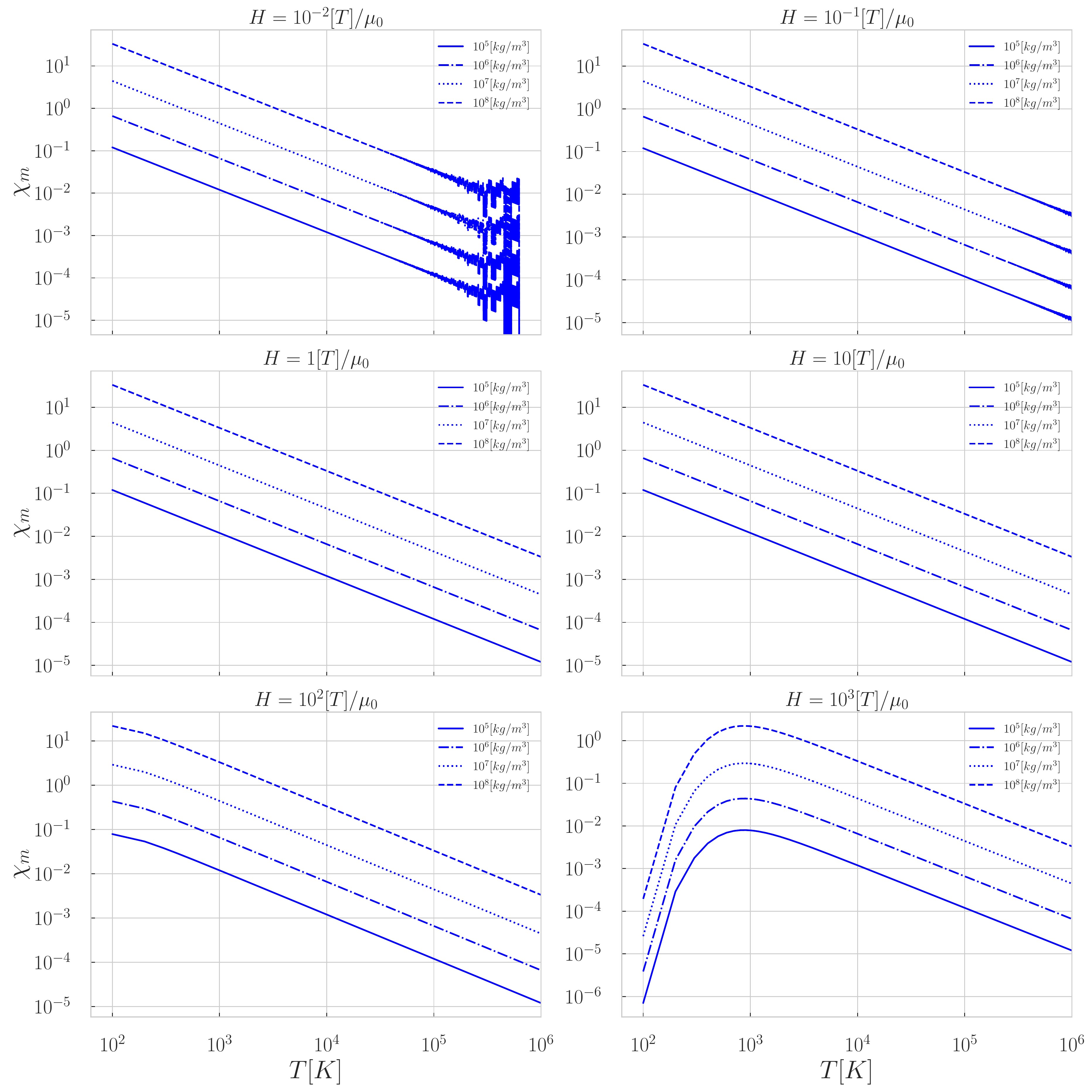}
    \caption{Magnetic susceptibility to paramagnetism as a function of temperature for different densities and magnetic fields in logarithmic scale. For magnetic fields between $10^{-2} [T]$ and $10^{2} [T]$, susceptibility decreases as temperature increases; however, for magnetic fields of $10^{3} [T]$, susceptibility increases as the temperature increases until it reaches a maximum from which it begins to decrease. In addition, it is noticeable that the magnetic field does not modify the magnitude of the magnetic susceptibility. One aspect to take into account is the noise that appears in the vicinity of the temperature $10^6 [K]$ for magnetic fields of $10^{-2} [T]$ and $10^{-1} [T]$, which is due to the nature of the Brillouin function: asymptotic nature of hyperbolic trigonometric functions makes not possible to calculate the derivative of magnetization in the vicinity of critic temperatures with precision. As the magnetic field increases, the range of admissible temperatures for calculating the magnetic susceptibility expands. This is evidenced by the fact that the noise is shifted to the right in the upper right panel, where it is greatly reduced, but still noticeable. On the rear panels, such noise is no longer perceived.}
    \label{fig:chi_T}
\end{figure*}

On the other hand, for paramagnetic materials, density, temperature, and magnetic field affect magnetic susceptibility.  In figure \eqref{fig:chi_T} we show the magnetic susceptibility for different density values as a function of temperature.
For accretion disks, temperatures for cool accretion in some astrophysical scenarios are in the range of $10^{2}[K] - 10^{4} [K]$, while hot gas can reach temperatures of $10^7[K] - 10^9 [K]$ \cite{cold_plasma}. For example, it is believed that the nucleus of the X-ray source Cygnus X-1 is a stellar-mass black hole surrounded by a gas with a temperature of $\sim 10^4[K]$. Furthermore, we show the influence of the magnetic field considering scales from $10^{-2}[T] - 10^{-1} [T]$. The magnetic field of these orders of magnitude may star the accretion process, for example, O-type stars feeding $SgrA*$ or in X-ray binaries systems \citep{mag_field}. We found that the magnetic field does not affect significantly the magnitude of the magnetic susceptibility in the range between $10^{-2}[T]$ and $10[T]$, but amplifies the range of temperatures by which it is possible to calculate the magnetic susceptibility. This is due to a critical value in the temperature that restricts the range of the magnetic susceptibility. This behavior is evidenced in the first row of the figure \eqref{fig:chi_T}. Nevertheless, if the magnetic field is stronger, for example in the range of $10^{2} [T]- 10^{3} [T]$, the magnetic field modifies the magnitude of the susceptibility. In the last row of figure \eqref{fig:chi_T} $\chi_m$ increases as $T$ increases too until a maximum value is reached, from which $\chi_m$ begins to decrease. In general, for low temperatures, it is possible to obtain susceptibilities in order of $10^{1} - 10^{-1}$, but for high temperatures and densities, we found orders of $10^{-1} - 10^{-2}$.

Magnetic properties of plasmas have been studied in quantum collisional and collisionless plasmas \citep{Latyshev} or in the propagation of waves quantum plasmas \citep{Safdar, Nauman} considering the Landau diamagnetism and Pauli paramagnetism applicable astrophysical scenarios like neutron stars, extragalactic jets or in auroral forms, where the acceleration of electrons is generated by magnetic fields. Expression for magnetic susceptibility in these cases is similar to equation \eqref{eq:chi_diam} where the relation between $H$ and $T$ is equivalent to equation \eqref{eq:alpha}. For this reason, it is believed that plasmas described by Landau diamagnetism and Pauli paramagnetism do not would exhibit a behavior much different concerning the model employed in this work.

\subsection{Geometrical considerations}
The Kerr space-time line element in Boyer-Lindquist coordinates, $\left\lbrace t, r, \theta,\phi \right\rbrace$, reads as follows
\begin{equation}
\begin{split}
    & ds^2 = -\left(1-\frac{2Mr}{\Sigma}\right)dt^2 + \frac{\Sigma}{\Delta}dr^2 + \Sigma d\theta^2 +
    \\
    & \left(r^2 + a^2 +  \frac{2Mra^2\sin^2\theta}{\Sigma} \right)\sin^2\theta
    d\phi^2 - \frac{4Mra\sin^2\theta}{\Sigma}dtd\phi,
    \end{split}
\end{equation}
with $a$ the dimensionless spin parameter and
\begin{equation}
\begin{split}
   &\Sigma = r^2 + a^2\cos^2(\theta),
   \\ 
   & \Delta = r^2-2Mr+a^2.
    \end{split}
\end{equation}

 On the other hand, it is useful to describe the torus structure in terms of the specific angular momentum, $l$, and the specific angular velocity, $\Omega$, defined as
\begin{equation}
\begin{split}
& l = -u_\phi/u_t = -\frac{g_{t\phi} + \Omega g_{\phi\phi} }{g_{tt}+\Omega g_{t\phi}},
\\
& \Omega =  u^\phi/u^t = -\frac{g_{t\phi}+lg_{tt}}{g_{\phi\phi}+lg_{t\phi}},
    \end{split}
\end{equation}
where $g_{\mu\nu}$ are the covariant components of the metric tensor. We define the center and cusp of the torus, as the places where the pressure gradient equals zero, so the specific angular momentum becomes Keplerian \citep{bardeen1972rotating},
\begin{equation}
l_K = \frac{M^{1/2}[r^2-2a(Mr)^{1/2}+a^2]}{r^{3/2}-2Mr^{1/2}+aM^{1/2}}, \label{eq:keplerian}
\end{equation}
thus, it is possible to calculate the positions on the equatorial plane for the center, $r_{\text{center}}$, and the cusp, $r_{\text{cusp}}$, solving \eqref{eq:keplerian} for a specific value of $l_K$ and $a$. Outside the event horizon, $r_{\text{center}}$ and $r_{\text{cusp}}$ are the largest and the smallest of the roots of \eqref{eq:keplerian}, respectively. 
\subsection{Magnetohydrodynamic structure of the torus}
In our disk model, we consider the test fluid approximation, namely, we ignore the gravitational effects generated by the disk. On the other hand, the total energy-momentum tensor is composed of the sum of the energy-momentum tensors for a perfect fluid, a magnetized fluid, and a fluid with magnetic polarization \citep{Oscar}
\begin{equation}
    \begin{split}
    T^{\mu\nu} = & \left[ w + b^2(1-\chi) \right]u^\mu u^\nu + 
    \\ 
    & \left[ p+\frac{1}{2}b^2(1-2\chi) \right]g^{\mu\nu}-(1-\chi)b^\mu b^\nu,
    \end{split}
\end{equation}
where $w$ is the specific enthalpy; $b^\gamma$ and $b$ are the components and magnitude of the magnetic field measured in the comovil frame; $u^\gamma$ are the components of the four-velocity of the disk, $p$ is the fluid pressure, and $g^{\mu\nu}$ are the contravariant components of the metric tensor. Furthermore, 
\begin{equation}
\chi = \chi_m/(1+\chi_m),
\end{equation}
 where $\chi_m$ is the magnetic susceptibility of the medium. In particular, for a magnetized, stationary, and axially-symmetric fluid around a Kerr black hole, the state variables do not depend on the coordinate time, $t$, or azimuthal angle, $\phi$. Besides, we assume that the fluid moves in circular orbits, such that $u^r$ = $u^\theta$ = 0. Furthermore, we consider that $b^r=b^\theta = 0$, which implies that the topology of the magnetic field is purely toroidal. 
 
 From the conservation of the energy-momentum tensor, $\nabla_\mu T^{\mu\nu} = 0$, it is possible to show that
\begin{equation}
   (\ln|u_t|)_{,i} - \frac{\Omega l_{,i}}{1-l\Omega} + \frac{p_{,i}}{w} - \frac{(\chi p_m)_{,i}}{w} + \frac{\left[ (1-\chi)\mathcal{L}p_m \right]_{,i}}{\mathcal{L}w} = 0, \label{eq:euler1}
\end{equation}
where the subscript $,i$ denotes the usual derivatives with respect to $r$, $\theta$, and $p_m = b^2/2$ is the magnetic pressure, and
  \begin{equation}
   \mathcal{L} = g_{t\phi}^2-g_{tt}g_{\phi\phi}.
\end{equation}
The first term in \eqref{eq:euler1} describes the gravitational interaction between the black hole and the torus, the second one refers to the centrifugal force due to the rotational motion of the disk, the third one is a force due to the pressure gradient, and the last terms are associated with magnetic forces. It is worth mentioning that if $\chi = 0$, \eqref{eq:euler1} reduces to Komissarov's case, where the torus is not magnetically polarized. 

In our study, we assume that the surfaces of $l$ and $\Omega$ constant coincide, namely $\Omega = \Omega(l)$; besides, the torus moves with constant specific angular momentum, $l_K = l_0$, and the fluid obeys the equation of state
\begin{equation}
p = K w^\kappa,
\end{equation}
with $K$ the polytropic constant and $\kappa$ the adiabatic index. Besides, assuming that $\chi$ can be written as a functional of $\mathcal{L}$, $\chi = \chi(\mathcal{L})$, the equation \eqref{eq:euler1} can be completely solved. Thus
\begin{equation}
 W - W_{\text{in}} + \frac{\kappa}{\kappa - 1}\frac{p}{w} + (1-2\chi)\frac{\eta}{\eta - 1}\frac{p_\text{m}}{w} = 0, \label{eq:solution}
 \end{equation}
where we define the effective potential, W, as follows
\begin{equation}
    W = \ln|u_t| = \ln|\mathcal{L}/\mathcal{A}|,
\end{equation}
with
\begin{equation}
\mathcal{A} = g_{t\phi} + 2lg_{t\phi}+l^2g_{tt}.
\end{equation}
Furthermore, $W_{\text{in}}$ is the effective potential in the inner edge of the disk and $\eta$ is an arbitrary constant. From \eqref{eq:solution} it is possible to calculate $p$ and $p_m$ at the disk center
\begin{equation}
\begin{split}
& p_\text{c} = w_\text{c}(W_{\text{in}} - W_\text{c}) \left( \frac{\kappa}{\kappa - 1} + \frac{\eta}{\eta - 1} \frac{1-2\chi_\text{c}}{\beta_\text{c}}  \right)^{-1}, 
\\
& p_\text{m$_\text{c}$} = \frac{p_\text{c}}{\beta_\text{c}}.
\end{split}
\end{equation}
where the subscript "c" refers to variables evaluated at the center of the disk on the equatorial plane. Now, assuming a magnetic susceptibility of the form
\begin{equation}
    \chi = \chi_0 + \chi_1\mathcal{L}^\alpha,
\end{equation}
 with $\chi_0$, $\chi_1$, and $\alpha$ constants, we can calculate the magnetic pressure
\begin{equation}
p_m = K_m \mathcal{L}^\lambda w^\eta f,
\end{equation} 
with
\begin{equation}
\begin{split}
& K_\text{m} = \frac{p_{\text{m}_\text{c}}}{\mathcal{L}_\text{c}^\lambda w_\text{c}^\eta f_\text{c}},
\quad
f = (1 - 2\chi)^{\frac{1-\eta}{2\alpha\left(1 - 2\chi_0\right)} - 1},
\\
& \lambda = \frac{1-\chi_{0}}{1-2 \chi_{0}}(\eta-1).
\end{split}
\end{equation}
Furthermore, physical variables such as the enthalpy, $w$, and the mass density, $\rho$, can be written as follows
\begin{equation}
\begin{split}
&\rho = w - \frac{\kappa p}{\kappa - 1},
\\ 
&w = \left( \frac{W_{\text{in}} - W} {\frac{\kappa}{\kappa - 1}K + \frac{\eta}{\eta - 1} (1-2\chi)K_\text{m}\mathcal{L}^\lambda f}  \right)^{\frac{1}{\eta-1}},
 \end{split}
\end{equation}
where
\begin{equation}
K = \frac{p_\text{c}}{w_\text{c}^\eta}.
\end{equation}
Finally, the components of the magnetic field are completely specified as
\begin{equation}
b^{\phi} = \sqrt{\frac{2p_m}{\mathcal{A}}}, \quad b^t = l_0 b^{\phi},
\end{equation}
In this model, the torus structure has the following free parameters: $l_0$, $\beta_{\text{c}}$, $w_\text{c}$, $\kappa$, $\eta$, $\alpha$, $\chi_m$, $\chi_1$, and $W_\text{in}$, which will be used as conditions for the structure and morphology of the torus in our simulations. 

\section{Radiative Transfer Equations}
\label{sec:RTE}

To get an intensity map of the radiation coming from the disk as a function of the torus parameters, we describe the radiative transfer through the covariant formulation  \citep{rybicki}
\begin{equation}
    \frac{d}{d\lambda}\left( \frac{I_\nu}{\nu^3} \right) = \left( \frac{j_\nu}{\nu^2} \right) - (\nu\alpha_\nu)\left( \frac{I_\nu}{\nu^3}\right), \label{eq:radtrans}
\end{equation}
where $I_{\nu}$, $j_{\nu}$, and $\alpha_{\nu}$ are the specific intensity, the emission coefficient, and the absorption coefficient, respectively, measured in the comovil frame, $\lambda$ is an affine parameter and the terms in parentheses are Lorentz invariants. The subscript $\nu$ indicates frequency dependence. To solve \eqref{eq:radtrans}, it is useful to rewrite the equation in terms of the optical depth, $\tau_\nu$, defined as
\begin{equation}
    \tau_\nu  = \int_{\lambda_0}^{\lambda}\nu\alpha_\nu d\lambda', \label{eq:tau}
\end{equation}
from which it is possible to classify the torus as optically thick, if $\tau_\nu = 0$, or optically thin, if $\tau_\nu > 0$. Usually, a minus sign appears if the integration is done backward along the path of photons inside the disk. Rewriting \eqref{eq:radtrans} in terms of $\tau_\nu$, it is possible to find an analytic solution to the differential equation, which reads as follows
\begin{equation}
     I_\nu(\tau_\nu)  = I_\nu(\tau_{\nu,0})\text{e}^{-\tau_\nu} + \int_{\tau_{\nu,0}}^{\tau_{\nu}}S_\nu(\tau'_\nu)\text{e}^{-(\tau'_\nu - \tau_{\nu,0})} d\tau'_\nu,
\end{equation}
where $S_\nu = j_\nu/\alpha_\nu$ is known as a source function, and if that is constant concerning $\tau_{\nu}$
\begin{equation}
     I_\nu(\tau_\nu) = I_\nu(0)\text{e}^{-\tau_\nu} + S_\nu(1 - \text{e}^{-\tau_\nu}). \label{eq:intensity}
\end{equation}

 Emission and absorption coefficients depend on the radiative process. In our case, we decided to study the effect of non-thermal synchrotron radiation as a power law with constant coefficients
\begin{equation}
    j_\nu \propto B^{(\gamma+1)/2}\nu^{(1-\gamma)/2}, 
    \quad
    \alpha_\nu \propto B^{(\gamma+2)/2}\nu^{-(\gamma+4)/2}, \label{eq:coefficents}
\end{equation}
where $\gamma = 2s+1$ with $s$ the spectral index, which is adjusted to the observations. The coefficients depend on the magnetic field, which is modified in magnetically polarized materials by the action of magnetic susceptibility. It is for this reason that synchrotron radiation is an interesting mechanism to prove the effect of magnetic polarization on the intensity map and observed flux. Through Lorentz invariants, it is possible to calculate the specific intensity measured by the distant observer. For it,
\begin{equation}
    I_{\nu_{\text{obs}}} = \left( \frac{\nu_{\text{obs}}}{\nu_{\text{em}}} \right)^3 I_{\nu_{\text{em}}}  = g^3 I_{\nu_{\text{em}}}, \label{eq:iobs}
\end{equation}
where the subscripts "obs" and "em" refer to intensity and frequency received by the observer and emitted by the disk, respectively, and $g = (1+z)^{-1}$  is the red-shift factor calculated as 
\begin{equation}
    g = \frac{\nu_{\text{obs}}}{\nu_{\text{em}}} = \frac{-p_{\mu}v^\mu|_{\lambda_\text{obs}}}{-p_{\mu}u^\mu|_\lambda},
\end{equation}
with $\left\lbrace v^\mu \right\rbrace = \left\lbrace-1,0,0,0\right\rbrace$ the four-velocity of the distant observer, and $p_\mu$ the components of four-momentum for photons. Without loss of generality, we set $p_{t_{\text{obs}}} = -E_{\text{obs}} = -1$ as a normalization condition. 

\section{Numerical Setup}
\label{sec:NS}

We carried out numerical simulations using \texttt{OSIRIS} ({\bf O}rbits and {\bf S}hadows {\bf I}n {\bf R}elativ{\bf I}stic {\bf S}pace-times) \citep{OSIRIS}, a code of our authorship based on the backward ray-tracing algorithm for stationary and axially-symmetric space-times. \texttt{OSIRIS} evolves null geodesics “backward in time” by solving the equations of motion in the Hamiltonian formalism. Our code works based on the image-plane model, an assumption where photons came from the observation screen in the direction of the black hole. Some authors employ this method to simulate shadows around black holes \citep{Johannsen_2013, doi:10.1142/S0218271816410212, disformal}, classifying the orbits of photons into two groups: those that reach the event horizon and those that escape to infinity. Every pixel on the image plane corresponds to an initial condition for each photon. Thus, when the code assigns a color to each pixel according to the initial conditions, we obtain an intensity map corresponding to the image of the radiation coming from the accretion disk around the black hole. 

The null-geodesic integration process is performed for two different angles, $\theta_0 = 45^\circ$ and $\theta_0 = 85^\circ$, assuming that the Minkowskian observer is located at $\left\lbrace t_0, r_0, \theta_0, \phi_0 \right\rbrace $ = $\left\lbrace 0, 1000, \theta_0, 0 \right\rbrace$. We define the surface of the tori, as the two-dimensional region where $\rho = 0.01$. Once the photon reaches this surface, the integration process of the radiative equation starts and keeps going while the photon remains inside the tori. For an optically thick disk $\tau_\nu$ = 0, implying that $I_{\nu_{\text{obs}}} \propto g^3$. For an optically thin disk, the algorithm calculates the value of $\lambda_0$ once the photon reaches the disk surface. Then, we use an Euler method to solve the equation \eqref{eq:tau} along the photon path inside the disk. Finally, the observation screen has a range $-12 \le x, y \le 12$ with a resolution of $1024\times 1024$ pixels for all simulations of the intensity map of the torus.
\section{Torus spectra}
\label{sec:KTMP}
\subsection{Exploring the space of parameters}
In particular, we study some combinations of parameters for paramagnetic and diamagnetic fluids about the case without magnetic polarization; whereby, it is interesting to compare the effects of magnetic polarization in the observed intensity and the morphology of the torus, which are observable characteristics. For this, we will consider a magnetically polarized disk with constant $\chi$ which implies that $\chi_1 = 0$, thereby $\chi = \chi_0 = \chi_m/(\chi_m+1)$. Thus, the magnetic susceptibility is constant which implies that $\chi_c$ and $f_c$ are also constants. 

In tables \eqref{tab:constants} and \eqref{tab:chis} we present the parameters that define the structure and morphology of the tori used in this paper. 
\\
\begin{table}
\centering
\begin{tabular}
{|c|c|c|c|c|c|}
\hline
$ a = 0.9$ & 
$l_0 = 2.8$ & 
$r_{\text{center}} = 4.622$ &
$r_{\text{cusp}} = 1.583$ \\
\hline
$W_{\text{in}} = -0.05$ & $W_{\text{c}} = -0.103$ & $\mathcal{L}_{\text{c}}=12.932$ &
$w_{\text{c}} = 1$ \\
\hline
$\kappa = 4/3$ & 
$\eta = 4/3$ & 
$\alpha = 1$ & 
$\chi_1 = 0$ \\ \hline
\end{tabular}
\caption{Set of constant parameters employed in all simulations. Some of these are inspired in the article of Komissarov.} \label{tab:constants}
\end{table}
\begin{table}
\centering
\begin{tabular}{cccccc}
& \multicolumn{5}{c}{}                                                              \\ \hline
\multicolumn{1}{|c|}{$\chi_m$}       & \multicolumn{1}{c|}{$-0.4$}   & \multicolumn{1}{c|}{$-0.2$}   & \multicolumn{1}{c|}{$0$} & \multicolumn{1}{c|}{$0.2$}   & \multicolumn{1}{c|}{$0.4$}   \\ \hline
\multicolumn{1}{|c|}{$\chi_\text{c}$}       & \multicolumn{1}{c|}{$-0.666$} & \multicolumn{1}{c|}{$-0.250$} & \multicolumn{1}{c|}{0}   & \multicolumn{1}{c|}{$0.166$} & \multicolumn{1}{c|}{$0.286$} \\ \hline
\multicolumn{1}{|c|}{$f_{\text{c}}$} & \multicolumn{1}{c|}{$0.403$}  & \multicolumn{1}{c|}{$0.637$}  & \multicolumn{1}{c|}{1}   & \multicolumn{1}{c|}{$1.660$} & \multicolumn{1}{c|}{$3.244$} \\ \hline
\end{tabular}
\caption{Values of $\chi_m$ and their corresponding values of $\chi_\text{c}$ and $f_\text{c}$.} \label{tab:chis}
\end{table}
Furthermore, we select constant emission and absorption coefficients as well as a spectral index $s = 0.75$ in the power-law synchrotron radiation model, which is motivated by observations of S-type radio sources \citep{Radio_astro}. This particular value for $s$ leads to $\gamma = 2.5$ in the coefficients of \eqref{eq:coefficents}.

\begin{figure*}
\centering
   \includegraphics[width=13.5 cm]{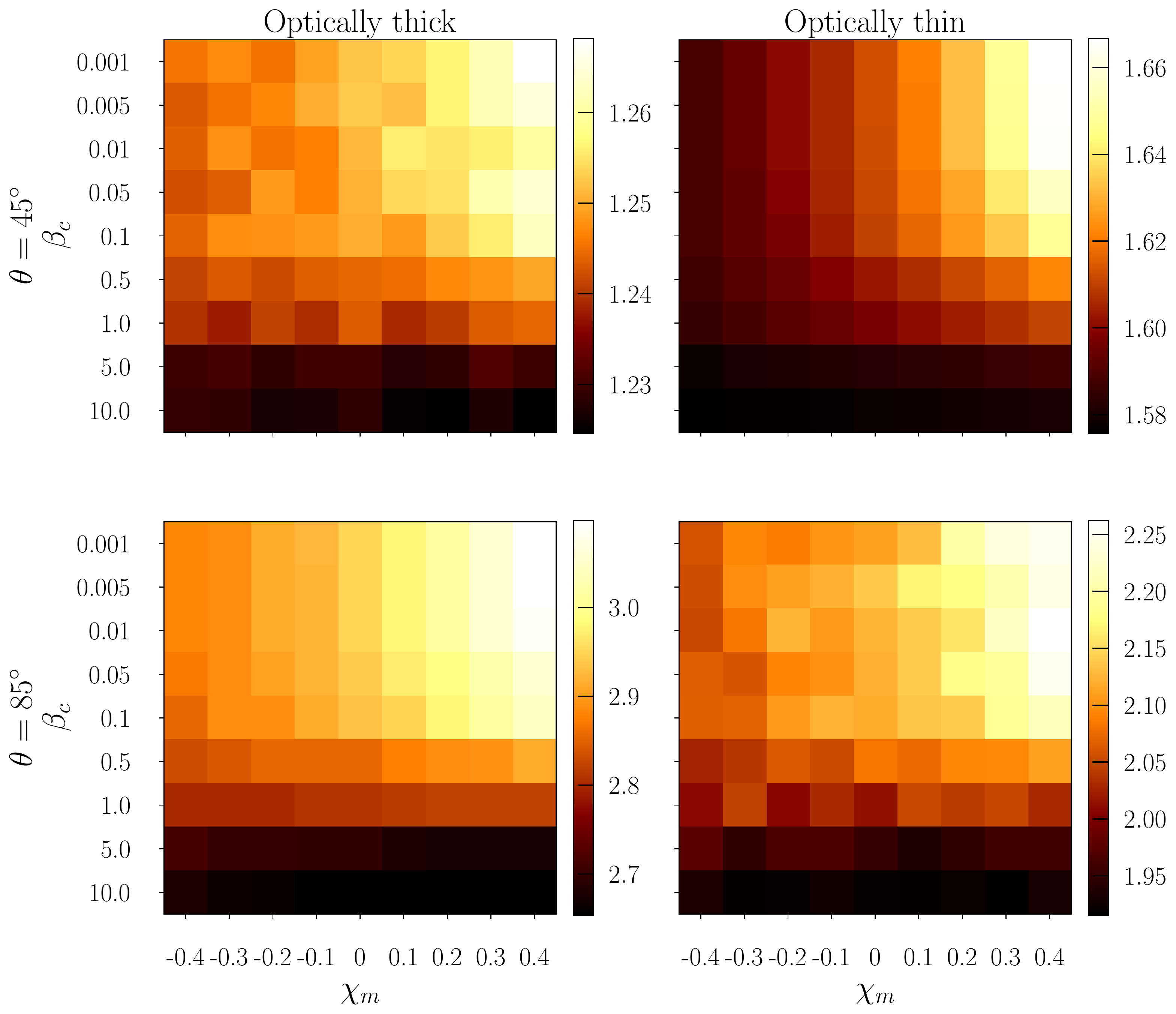}
    \caption{Maximum values of intensity emitted by an optically thick (left column) and optically thin (right column) torus around a Kerr black hole with dimensionless rotation parameter a = 0.9 seen from observation angles $\theta = 45^\circ$ (top row) and $\theta = 85^\circ$ (bottom row). It is clear that the effects of magnetic polarization are more relevant as $\beta_{\text{c}}$ decreases, and for paramagnetic disks, the tendency is to emit with higher intensity than in the case of diamagnetic disks. Besides, optically thin disks emit more intensely than optically thick disks if the angle of observation is $\theta_0 = 45^\circ$; nevertheless, if $\theta_0 = 85^\circ$ a more intensification emission comes from the optically thin disk.}
    \label{fig:sp}
\end{figure*}

On the other hand, the effects of the magnetic polarization should depend on the "beta-plasma" parameter, $\beta_c$, because it indicates the relative importance of the magnetic interactions as compared to the hydrodynamic forces. Based on this, we carried out a parameter study to establish which combination of these parameters allows the best visualization of the magnetic polarization effects. In figure \eqref{fig:sp} we show the maximum value of intensity measured for different combinations of $\chi_m$ and $\beta_{\text{c}}$, for both optically thick and optically thin disk viewed from $\theta_0 = 45^\circ$ and $\theta_0 = 85^\circ$. In this figure, it is possible to appreciate that for high values of beta-plasma (disk dominated by fluid pressure) the effects of magnetic polarization are practically negligible; in addition, for this range of values, the minimum intensity emitted is obtained in the four cases. On the other hand, for low values of beta-plasma (disk dominated by magnetic pressure) the effects of magnetic susceptibility become relevant, showing a tendency towards paramagnetic disks in terms of maximum values of emitted intensity. Likewise, it is possible to appreciate that these values also depend on the observation angle and the optical penetration of the photons into the disk. For an observation angle of $45^\circ$ the maximum value of intensity is achieved when the disk is optically thin ($I_{max} = 1.66$), while from an inclination of $85^\circ $ the maximum intensity emitted occurs when the disk is optically thick ($I_{max} = 3.0$). In general, the emission is more intense for 85$^\circ$ in both types of disks compared to the inclination of 45$^\circ$. 

\subsection{Emission-line profiles}

\begin{figure*}
\centering
   \includegraphics[width=17cm]{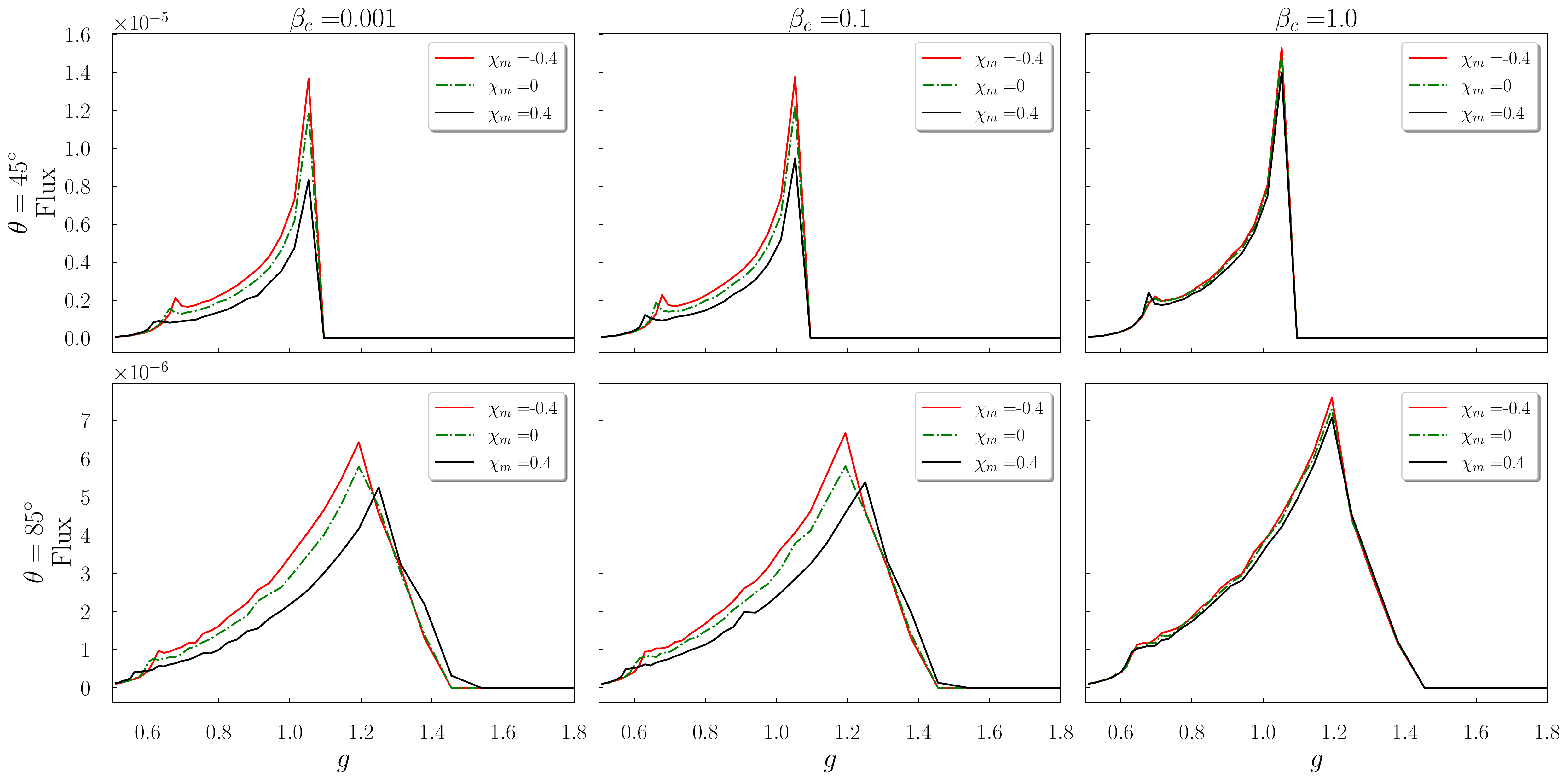}
    \caption{Flux spectra from an optically thick torus with $\beta=1.0$, $0.1$ , $0.001$ and for values of susceptibility $\chi_m = -0.4$, $0$, $0.4$ viewed form angles $\theta = 45^\circ$ (top) and $\theta = 85^\circ$ (bottom).}
    \label{fig:flux_opaque}
\end{figure*}

\begin{figure*}
\centering
   \includegraphics[width=17cm]{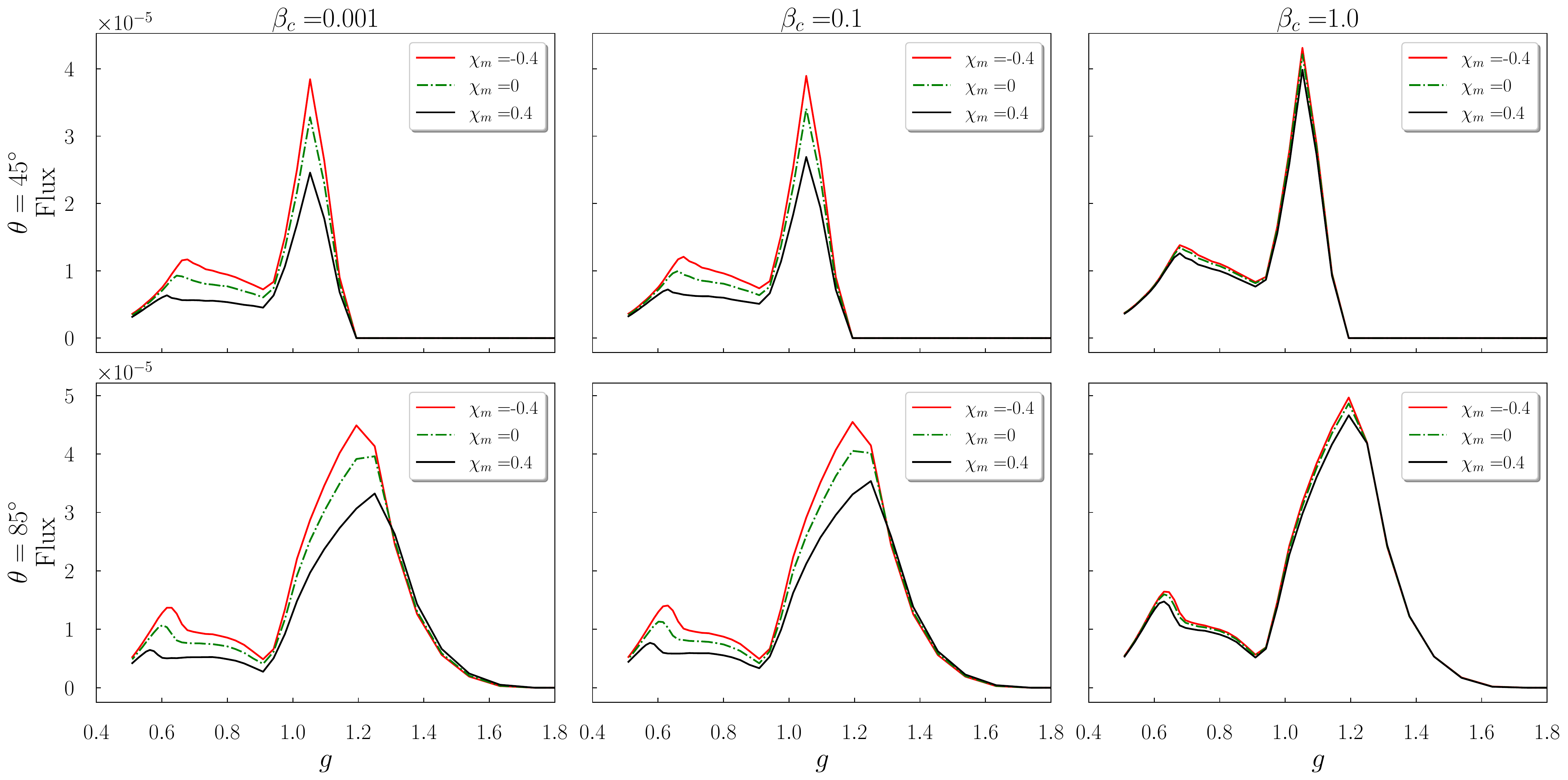}
    \caption{Flux spectra from an optically thin torus with $\beta=1.0$, $0.1$ , $0.001$ and for values of susceptibility $\chi_m = -0.4$, $0$, $0.4$ viewed form angles $\theta = 45^\circ$ (top) and $\theta = 85^\circ$ (bottom).}
    \label{fig:flux_trans}
\end{figure*}

Next, based on the previous results we decided to use the values at which the effects of magnetic susceptibility are most appreciable. For $\beta_c = 0.001, 0.1$ and $1.0$, and $\chi_c = -0.4, 0.0$ and $0.4$, the emission spectra were calculated by computing the flux from the tori, both optically thick and optically thin, knowing that flux can be expressed as
\begin{equation}
F_\nu = \int I_\nu d\Omega,
\end{equation}
where $d\Omega$ is the element of solid angle. Due to how \texttt{OSIRIS} is built under the image plane model, the solid angle element can be written as $d\Omega = dxdy/r_0^2$ (following a procedure similar to that of \cite{10.1111/j.1365-2966.2007.11855.x}), being completely specified when selecting the resolution of the simulations. We defined a frequency grid with a resolution $d\nu = 0.0375$ and with $\nu_i = 0.5$ and $\nu_f = 2.0$ as the minimum and maximum frequencies measured inside the torus, respectively. Then, the flux calculation is reduced by adding the intensity contributions for each frequency. We computed the frequency at each point in space and assumed that a photon was emitted with a specific frequency if the difference $|\nu_{\text{mesh}} - \nu_{\text{emitted}}|$ was less than a tolerance $\delta_\nu = 10^{-2}$. In that case, that intensity is considered a contributor to the total flux for that particular frequency. In figures  \eqref{fig:flux_opaque} and \eqref{fig:flux_trans} it is possible to appreciate the fluxes for the optically thick and optically thin torus, for different values of $\beta_c$, $\chi_c$, and $\theta_0$ as a function of the $g$ parameter.

First of all, the flux emitted by the paramagnetic torus is less than the flux emitted by the Komissarov and diamagnetic tori. This makes sense based on Pimentel's results, where paramagnetic tori are found to be more compact relative to the others. In the case of optically thick tori (figure \ref{fig:flux_opaque}), it is natural to think that the number of photons emanating from a paramagnetic torus surface is less than for a diamagnetic torus (less compact), due to covering less surface area and contributing less to the sum of intensities. Thus, a paramagnetic torus leads to a lower flux than a diamagnetic torus. In the case of optically thin tori (figure \ref{fig:flux_trans}), the optical path followed by the photons inside the paramagnetic torus is shorter (because of the mentioned above concerning surfaces), and therefore the contribution of the intensities, in this case, is lower in contrast to diamagnetic tori, where the photon travels more distance inside the torus. 

In all cases the effect of magnetic susceptibility decreases with increasing beta-plasma; however, it is remarkable that the maximum flux peak occurs for higher values of beta-plasma. This behavior can be explained due to the synchrotron radiation model that we used. From the coefficients shown in equation \eqref{eq:coefficents} it can be seen that the magnitude of the magnetic field is written as a power law of the spectral index. However, the source function, $S$, defined as the ratio between the emission and absorption coefficients, does not depend on that index and also results in the magnitude of the magnetic field always decreasing as $B^{-1/2}$. In this sense, as the magnetic field increases the contribution of the source function to the total intensity is less, which can be seen in equation \eqref{eq:intensity}. Based on the above, as the beta-plasma is smaller, it implies that the magnetic pressure is getting larger, which leads to a larger magnetic field. Thus, it is consistent that the flux decreases as beta-plasma also decreases, and increases as beta-plasma increases too. 

On the other hand, the angle of inclination also plays a role in the flux maximum. For $\theta_0 = 45^\circ$ and opaque torus (top of figure \ref{fig:flux_opaque}), we found that emission lines are larger than emission lines for opaque torus at 85$^\circ$ (bottom of figure \ref{fig:flux_opaque}). This behavior can be explained because the self-eclipsing at $85^\circ$ is higher than at $45^\circ$. Thus, if the observer is closer to the equatorial plane the observable region of the torus will be smaller and therefore the observed flux will be smaller also. Another interesting feature is that the emission profiles at $45^\circ$ for the torus (geometrically thick disk) are similar to emission profiles for a geometrically thin disk \citep{Schnittman_2004}, where we can observe two peaks: a smaller one to the left which corresponds to red-shifted emission that comes from material that is moving away from the observer, and a larger one to the right which corresponds to blue-shifted emission that comes from material approaching the observer \citep{fabian}. At $85^\circ$ the red-shift peak is considerably less than the blue-shift peak due to the self-eclipsing mentioned above, leading to a broad line that looks like a single horn. Besides, it is interesting how the paramagnetic torus at this inclination is blue-shifted in comparison with the other two cases. On the other hand, changing the viewing angle occurs a blue shift of the peaks, which is consistent with the results of \cite{JOVANOVIC201237}. 

Now, for both $45^\circ$ and $85^\circ$ inclinations for optically thin torus (figure \ref{fig:flux_trans}) the behavior is similar to previous cases, but with the main difference that the order of magnitude of the flux is the same for both, being higher for translucent torus flux than for both opaque cases. This is clear because the optical path followed by the photons inside of the optically thin torus is greater than in comparison with the optically thick torus, adding more intensities to the contribution of the total flux, keeping in mind that there are high-order lensed photons that orbit more than once the black hole, which contributes substantially to the flux. Also, in all the cases, the blue-shift peak is more noticeable, being a remarkable difference mainly for an inclination of $85^\circ$ when we compare the bottom rows of figures \eqref{fig:flux_opaque} and \eqref{fig:flux_trans}.

\subsection{Intensity map}

Finally, the results of our simulations for the intensity map from optically thick and optically thin emitting tori under the model of synchrotron radiation around a Kerr black hole are shown below. In figures \eqref{fig:torus_45} and \eqref{fig:torus_85} it is possible to see the photons coming from the disk that passes marginally near the event horizon in the last circular orbit and that manage to escape from the gravitational attraction, which forms or delimits the shadow of the black hole.

\begin{figure*}
   \includegraphics[width=17cm]{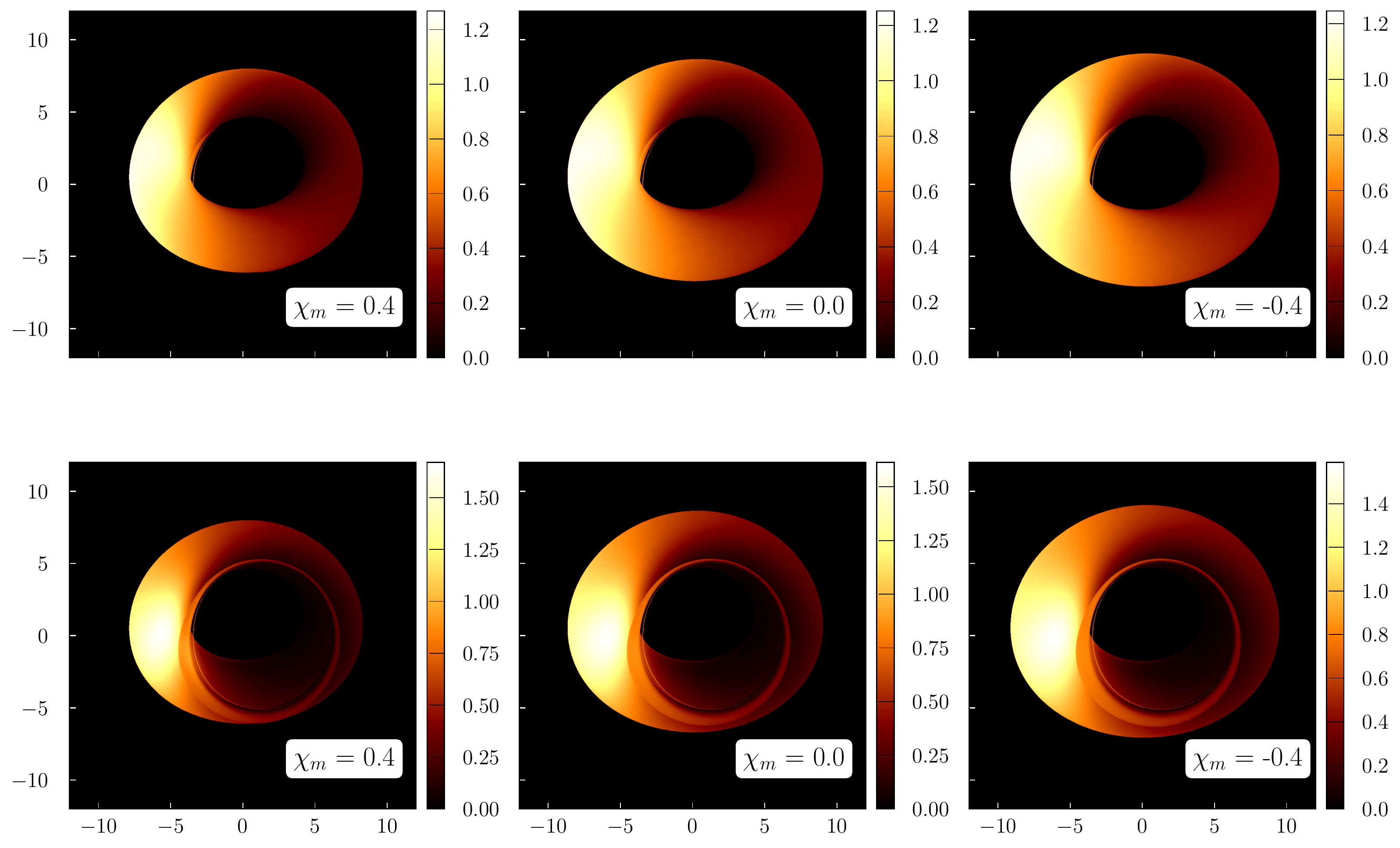}
    \caption{Magnetized torus optically thick (top panel) and optically thin (bottom panel) around a Kerr black hole with dimensionless spin parameter $a = 0.9$ view from $\theta = 45^\circ$  for different values of magnetic susceptibility $\beta = 0.001$.}
    \label{fig:torus_45}
\end{figure*}

\begin{figure*}
   \includegraphics[width=17cm]{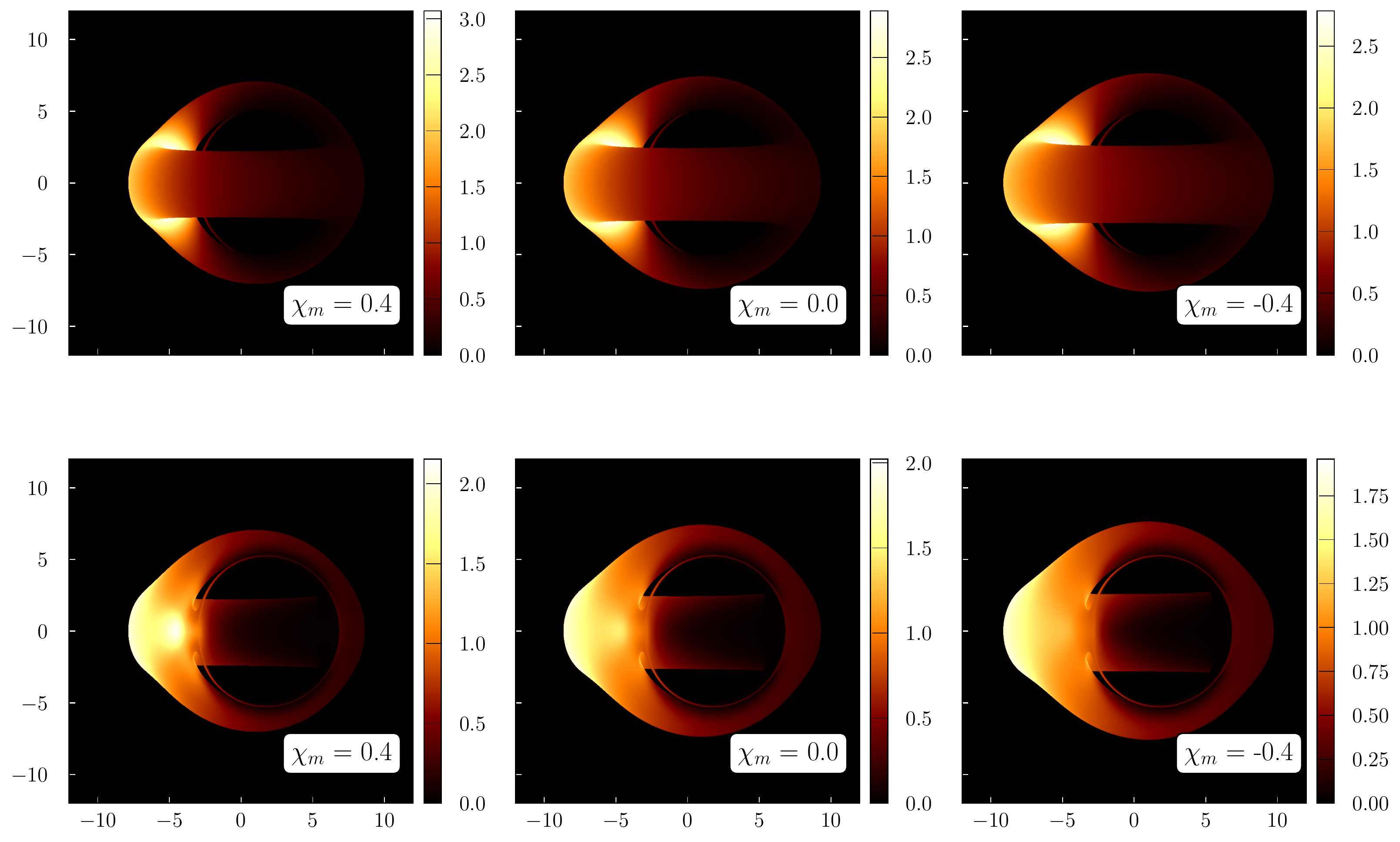}
    \caption{Magnetized torus optically thick (top panel) and optically thin (bottom panel) around a Kerr black hole with dimensionless spin parameter $a = 0.9$ view from $\theta = 85^\circ$  for different values of magnetic susceptibility $\beta = 0.001$.}
    \label{fig:torus_85}
\end{figure*}

Optically thick disks exhibit self-eclipsing because the radiation is assumed that comes only from photons at the surface. Photons inside the disk are absorbed and emitted in such a way that they do not reach the surface; in this sense, the image of the optically thick disks corresponds to the surface of the disk, and according to the equation \eqref{eq:iobs}, the map of intensities corresponds only to the factor $g$. For optically thin disks, it is possible to see the entire shadow of the black hole through the torus, in addition to being more clearly visible on the rear part of the torus, which in principle should be hidden from the observer. This effect, purposely caused by the deflection of light due to the strong gravitational field of the black hole, is more noticeable as viewing angles increase. 

On the other hand, as is expected from \cite{Oscar} results, it is evident that the paramagnetic tori (on the left) is more compact than the diamagnetic ones (on the right) compared to the \cite{komissarov} case (in the middle). This effect is reflected in the intensity profile, showing that there is a more concentrated region where the maximum intensity is located. This region lies to the left of the disk, corresponding to the direction of the spin of the black hole, and is more concentrated for the paramagnetic torus and spreads out further along the disk for the diamagnetic torus. Viewed from an 85$^\circ$ angle, the torus shape (bright bulge) is clear in the translucent paramagnetic disk, fading as the disk becomes more diamagnetic.

\section{Discussion \& Conclusions} \label{sec:conclusions}

In this paper, we simulate the intensity map and the emission-line profiles of the radiation coming from a torus around a Kerr black hole. We show the change of the magnetic susceptibility as a function of density, temperature, and magnetic field for both diamagnetic and paramagnetic cases. For diamagnetic plasmas, the susceptibility only depends on the density and exhibits a linear behavior. For paramagnetic plasmas, in particular, magnetic susceptibility decreases as temperature increases in the range of magnetic field between $10^{-2} [T] - 10[T]$, while for $10^{2} [T] - 10^{3}[T]$ for low temperatures magnetic susceptibility increases as temperature increases too, and returns to its previous behavior as temperature still increasing. For some configurations of density, temperatures, and magnetic field the magnetic susceptibility is of the order $10^{-3} - 10$, corresponding to values used in this work. Our torus is composed of a stationary and axially-symmetric fluid with arbitrary magnetic polarization and presents a toroidal magnetic field. Besides, we carry out a systematic study of the observed specific intensity and the observed flux as a function of the magnetic susceptibility, $\chi$, and the degree of magnetization, $\beta_{\text{c}}$. We found that assuming power-law synchrotron radiation with constant coefficients as the emission mechanism, the effects of magnetic polarization are negligible if the disk is dominated by the hydrostatic pressure, namely, $\beta_{\text{c}} > 1$. Moreover, if the disk is paramagnetic and dominated by the magnetic pressure, $\beta_{\text{c}} < 1$, the intensity reaches higher peaks for this configuration. 

On the other hand, the flux decreases as the degree of magnetization increases, which is observed in the emission-line profiles. This behavior is consistent with the fact that the source function decreases as $B^{-1/2}$. Furthermore, from these emission profiles is clear that diamagnetic fluid emits a higher flux because diamagnetic torus is less compact than paramagnetic ones. Therefore, photons have more optical paths to travel inside the disk and grow up the contribution to the total observed flux. It is highlighted that although the increase in the magnetic field decreases the maximum value of the flux, it does not modify the frequency at which this maximum of each of the fluxes occurs.

Furthermore, our simulations are consistent with the general results of torus spectra around the black hole. For optically thick disks, we found that for a low angle of observation the observed flux is similar to the observed flux for a geometrically thin and optically thick disk. If the observed angle is near the equatorial plane, the observed flux decreases and the peak of the maximum is blue-shifted. For optically thin disks we found that in general, the observed flux is higher than the measured in the case of optically thick disks. In this case, the blue and red horns are more visible even for observations near the equatorial plane. However, the emission lines are higher for high-angle inclination than low angle inclination. 

Another interesting feature is that magnetic polarization changes the observed flux, decreasing its maximum value as the material becomes more paramagnetic. Besides, magnetic susceptibility shifts the frequency at occurs the maximum flux: for paramagnetic materials, this maximum has a blue shift in the emission lines both optically thick and thin disk, which is more noticeable for $\theta_0 = 85^\circ$. These frequency shifts could be an observable characteristic that shows the degree of magnetic polarization of the fluid that conforms to the accretion disk. In this way, the effects of magnetic polarization would not be limited to a renormalization of the magnitude of the magnetic field but would have a potential relevance when comparing the observations with the numerical simulations.

\section*{Acknowledgements}

 F.D.L-C was supported by the Vicerrectoría de Investigación y Extensión - Universidad Industrial de Santander, under Grant No. 3703.


\end{document}